\documentclass[fleqn,10pt]{wlscirep}
\usepackage{graphics, float}
\title{Divisibility patterns of natural numbers on a complex network}

\author[1]{Snehal M. Shekatkar}
\author[1]{Chandrasheel Bhagwat}
\author[1,*]{G. Ambika}
\affil[1]{Indian Institute of Science Education and Research, Pune, 411008, India}

\affil[*]{g.ambika@iiserpune.ac.in}

\begin{abstract}
Investigation of divisibility properties of natural numbers is one of the most important themes in the theory of numbers. Various tools have been developed over the centuries to discover and study the various patterns in the sequence of natural numbers in the context of divisibility. In the present paper, we study the divisibility of natural numbers using the framework of a growing complex network. In particular, using tools from the field of statistical inference, we show that the network is scale-free but has a non-stationary degree distribution. Along with this, we report a new kind of similarity pattern for the local clustering, which we call ``stretching similarity", in this network. We also show that the various characteristics like average degree, global clustering coefficient and assortativity coefficient of the network vary smoothly with the size of the network. Using analytical arguments we estimate the asymptotic behavior of global clustering and average degree which is validated using numerical analysis.
\end{abstract}
\begin{document}

\flushbottom
\maketitle
\thispagestyle{empty}

\section*{Introduction}

The study of complex networks has become a very important part of many disciplines like information \cite{Broder_2000},technology \cite{Faloutsos_1999}, social sciences \cite{Amaral_2000}, ecology \cite{Huxham_1996} and biology \cite{Jeong_2000, Jeong_2001, White_1986}. The characterization of structure of real networks is an indispensable part of this study. Despite being random, real networks show certain statistical properties which set them apart from their completely random mathematical counterparts. This hints towards underlying organizing principles which shape the structures of real networks \cite{Networks_review}. In particular, many real networks are scale-free which means that the distribution of degrees of their nodes follows a power law \cite{Networks_review,Newman_book}. The density of triangles in the network is another important characteristic of networks measured using a quantity called clustering coefficient. Empirical studies show that the real networks are highly clustered as compared to completely random mathematical models like Erdos-Renyi graph \cite{Networks_review,Newman_book}. 

In the present paper, we report an analysis for a particular deterministic network that resembles real networks in many aspects. This network consists of natural numbers $1,2,3,\cdots$ as nodes and if a given number divides another, then their corresponding nodes are connected by an undirected link. The network thus constructed, though deterministic, can be studied on an equal footing with the other random networks because of the irregular distribution of primes which makes divisibility relations themselves irregular. It is helpful to view this network as a growing network where nodes are added one at a time. A similar network with nodes as composite numbers has already been studied \cite{Zhou_2005}. Also, a directed network of natural numbers based on the divisibility which includes only the multiples in the pattern has been reported by Ding-hua et al \cite{Ding_2010}. A bipartite structure separating composite and prime numbers with weighted links between them based on divisibility has been analyzed by Garc{\'i}a-P{\'e}rez et al \cite{Perez_2014}. 

In the present work we consider a more general set up where we put all the natural numbers on a complex network with their divisibility relations as the underlying deterministic rule of connections. Here the network is undirected with links to both divisors and multiples. Using tools from statistical inference, we confirm that this network is scale-free and show that average degree, global clustering coefficient and assortativity coefficient vary smoothly with the size of the network. This is surprising in view of the fact that distribution of primes is quite irregular in the sequence of natural numbers. We provide analytical results for the asymptotic behavior of average degree and global clustering coefficient for this network. In particular, we show that the global clustering coefficient of this network decays to zero whereas average degree increases logarithmically. We also report an interesting and novel similarity exhibited by local clustering coefficients of nodes in this network which we call ``stretching similarity". 

The remaining paper is organized as follows: In the next section we describe the construction of the network and show that the network is scale-free. We then describe the existence of stretching similarity in this network. Finally we show the behavior of average degree, global clustering coefficient and assortativity coefficient as a function of size of the network and analytically obtain the asymptotic trends for average degree and clustering.

\section*{Results}
\label{Results}
\begin{figure}[ht]
\begin{center}
\includegraphics[width=1\columnwidth]{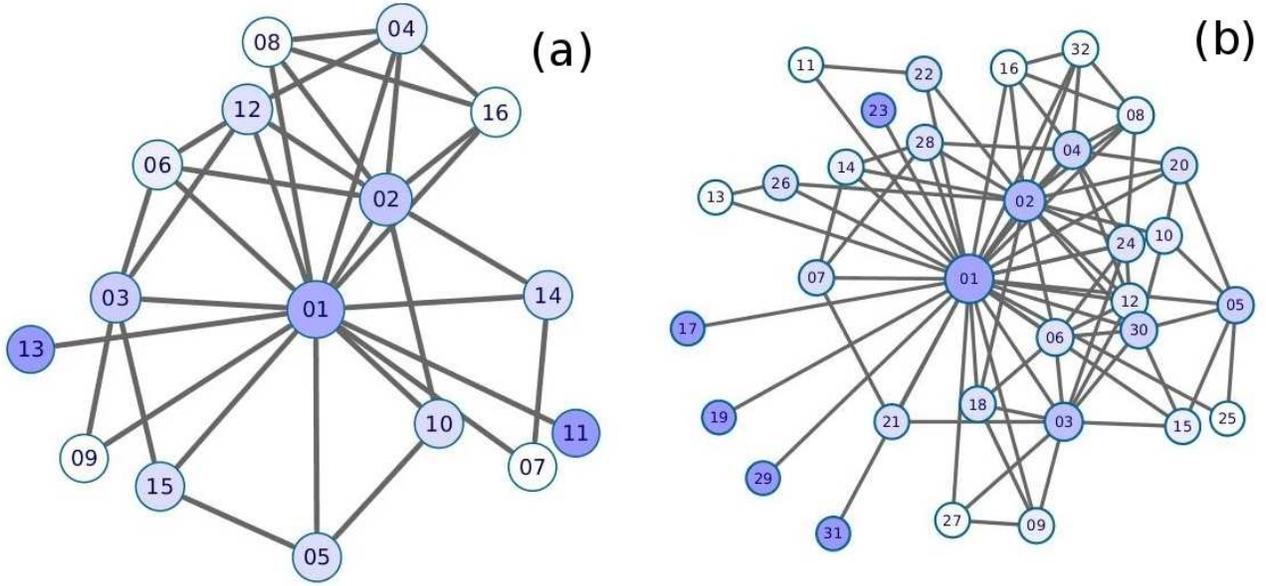}
\caption{\label{network_pic} \textbf{Network of natural numbers with two different sizes.} (a) $t=16$ nodes and (b) $t=32$ nodes. In each panel, the size of each node is proportional to its degree and color of each node is graded according to its clustering coefficient with more white nodes as nodes with higher value of local clustering. }
\end{center}
\end{figure}

\begin{figure}[ht]
\begin{center}
\includegraphics[width=1\columnwidth]{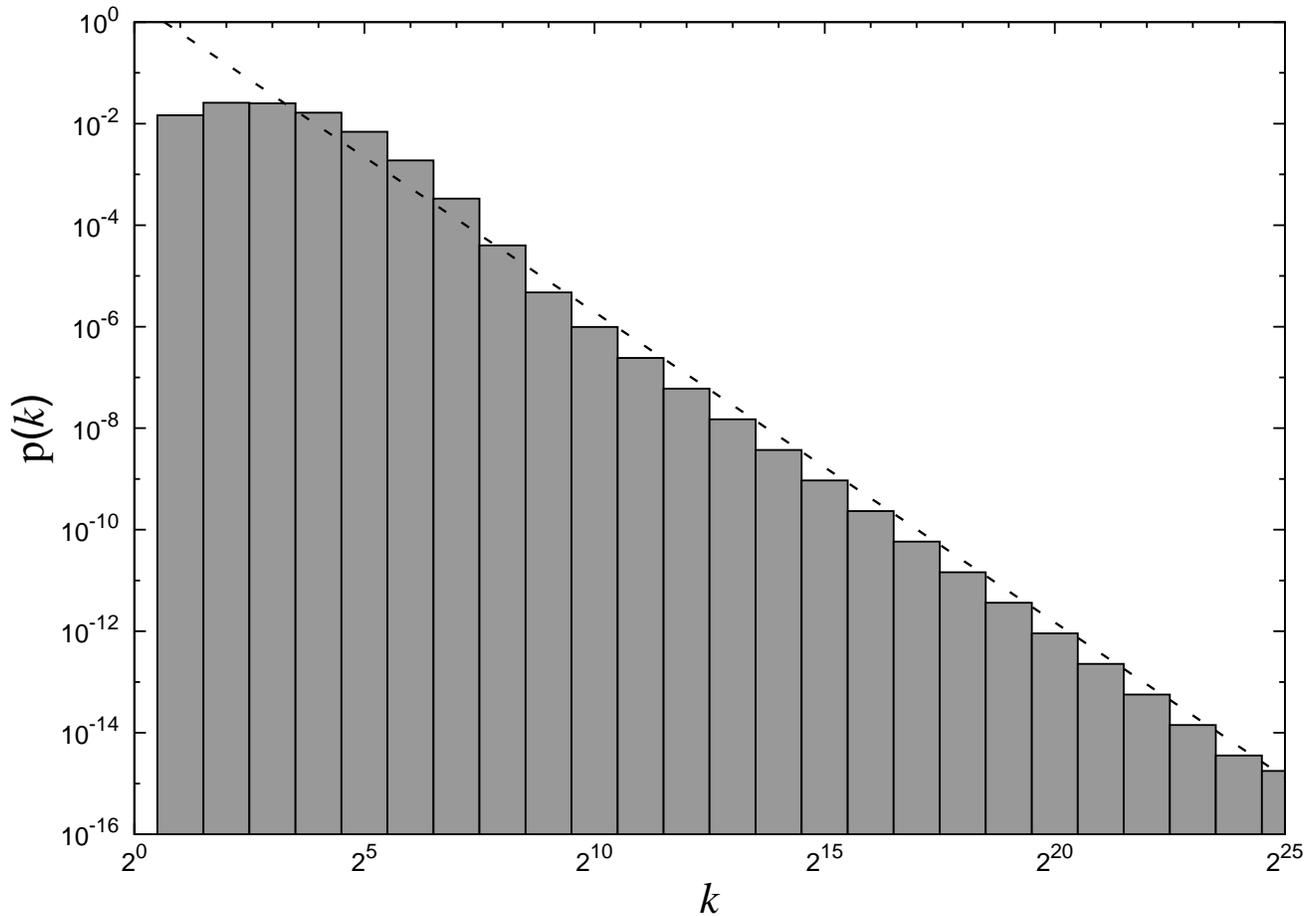}
\caption{\label{deg_distri} \textbf{Degree distribution of network of natural numbers with logarithmic binning.} Sizes of successive bins are equal to successive positive powers of $2$ and count in each bin is normalized by dividing by a bin width. The dotted line in the graph has slope $\alpha = -2$ and it is calculated using the \textit{method of maximum likelihood} \cite{Clauset_2009}. The existence of the underlying power law is established by calculating $p$-value using Kolmogorov-Smirnov statistic for smaller sizes of the same network (see Methods).}
\end{center}
\end{figure}

\textbf{Construction of the network and its scaling properties.} The nodes of the present network are natural numbers $1,2,3,\cdots$ and there is a link between two nodes if either divides the other. We avoid self-links and all the links are undirected. Since the sequence of natural numbers has natural ordering, it is helpful to view this network as a growing network with the addition of a new node at each discrete time as follows:
\begin{enumerate}
  \item{At time $t=1$ network starts with a single node $n=1$ and at every time $t$, a node with the number $n=t$ is added to the network.}
  \item{This node connects to all the existing nodes whose numbers divide it.}
\end{enumerate}
The network thus constructed is shown in Fig.~\ref{network_pic} at two different times $t=16$ and $t=32$ which would correspond to networks of size $N = 16$ and $N = 32$ respectively.
To find the distribution of degrees of this network, we grow the network till the size reaches $N = 2^{25} = 3,35,54,432$. The resulting distribution shown in Fig.~\ref{deg_distri} seems to follow a power law ($p(k)\sim k^{-\alpha}$) asymptotically. Using the method of maximum likelihood we find that the scaling-index $\alpha \sim 2$. We establish the existence of power-law in the distribution (and hence the fact that this network is scale-free) using the approach described in Clauset et al \cite{Clauset_2009} (see Methods).
We also study the scaling behavior of the local clustering coefficient with degree. The local clustering of a node in the network is defined as the fraction of number of edges that are present among its neighbors. For node $i$ with degree $k_{i}$ this can be written as \cite{Ravasz_2003}:
\begin{equation}
c_{i} = \frac{E_{i}}{^{k}C_{2}}
\end{equation}
where $E_{i}$ is the actual number of edges among the neighbors of node $i$.

\begin{figure}[ht]
\begin{center}
\includegraphics[width=1\columnwidth]{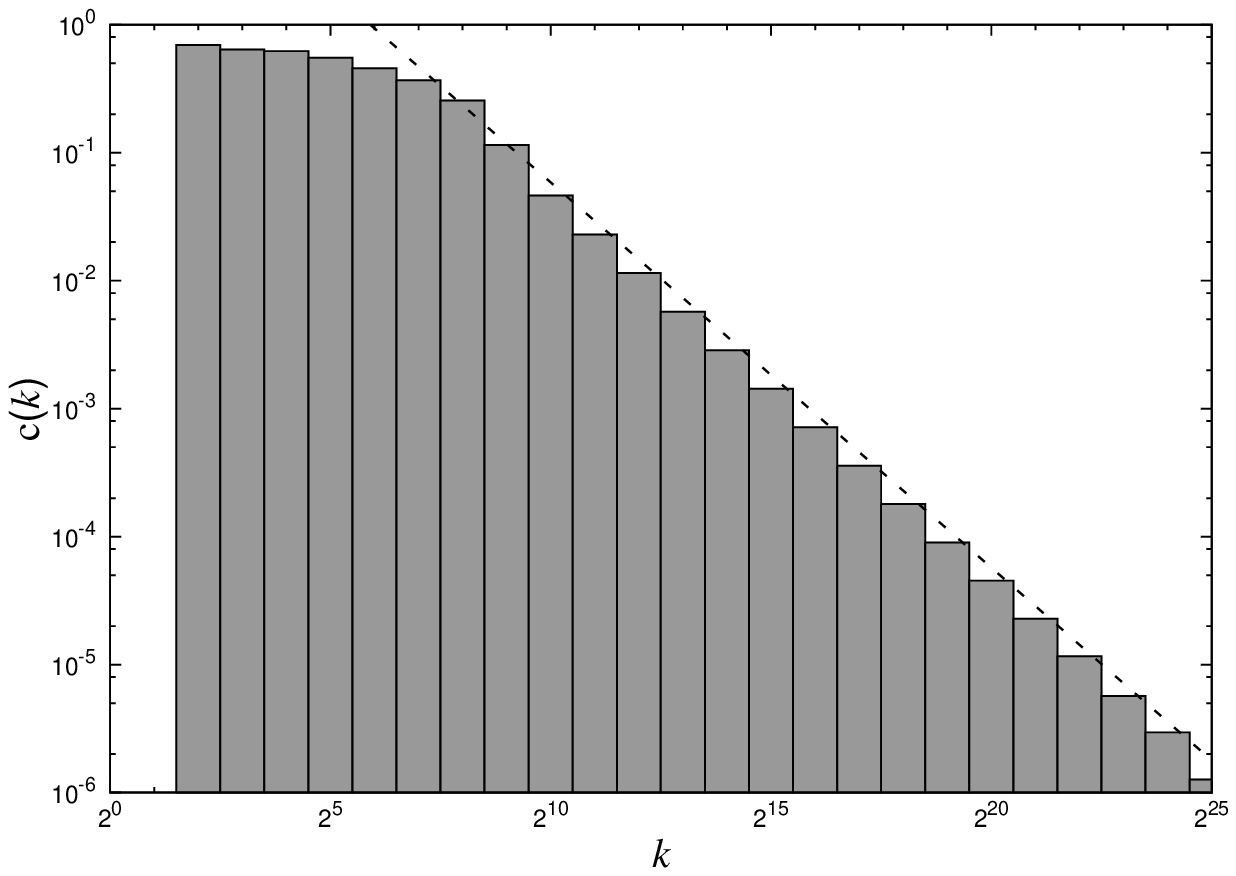}
\caption{\label{cluster_vs_deg} \textbf{Dependence of local clustering coefficient on degree.} The plot is created using a logarithmic binning. Asymptotically, the local clustering is seen to follow a power law with exponent $\sim 1$. }
\end{center}
\end{figure}

\begin{figure}[ht]
\begin{center}
\includegraphics[width=1\columnwidth]{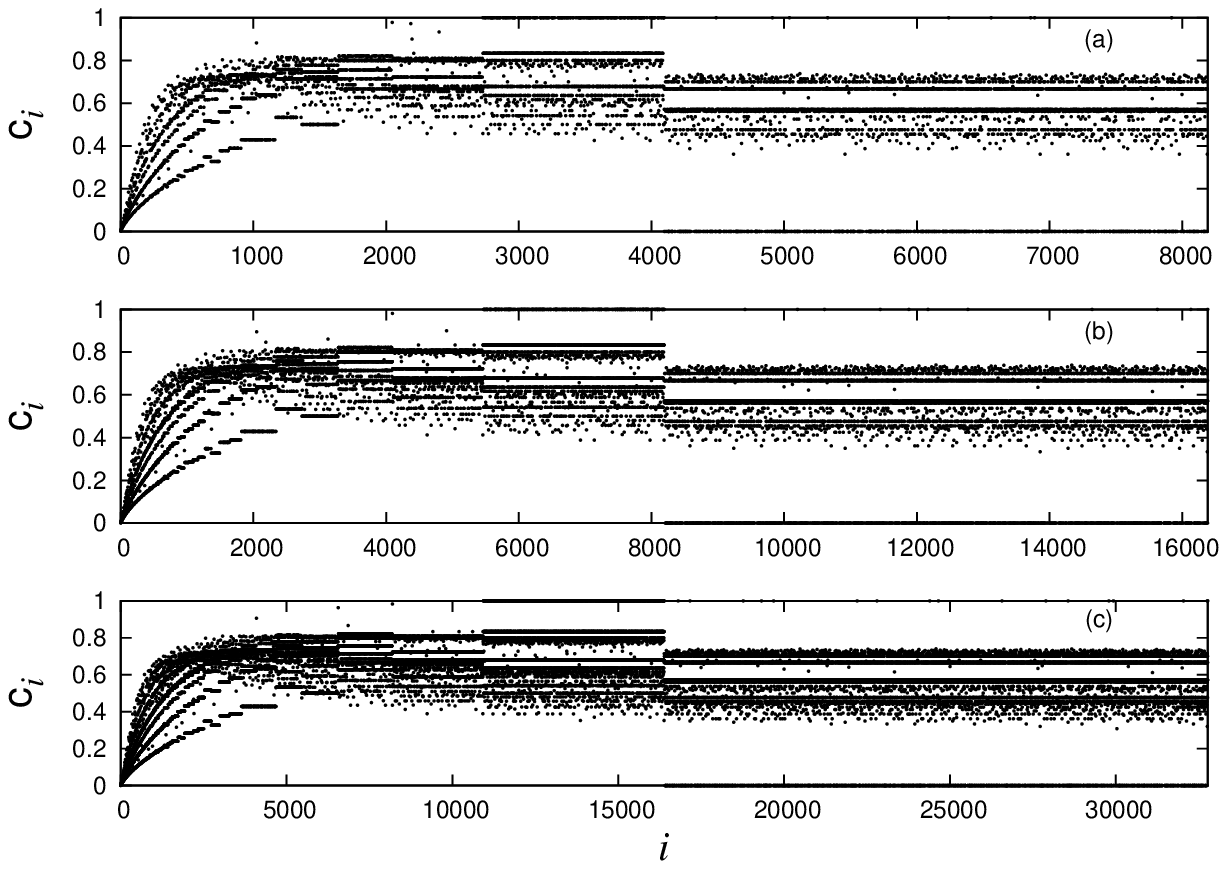}
\caption{\label{cluster_stretch} \textbf{Local clustering coefficient as a function of node index for three different sizes of network.} (a) $N=2^{13}$, (b) $N=2^{14}$ and (c) $N=2^{15}$. In any local region of the plot, the values $c_{i}$ seem to be scattered irregularly. However, with the increase in the network size, the whole pattern is stretched on a global scale. We call this similarity as ``stretching similarity".}
\end{center}
\end{figure}

In Fig.~\ref{cluster_vs_deg} we show the dependence of local clustering coefficient of nodes in the network on the degree. It can be seen that the asymptotic behavior is compatible with a power law with exponent $1$. This behavior is similar to one that is usually observed in real networks \cite{Ravasz_2003}. 
\textbf{Stretching similarity of local clustering.} We now discuss an interesting behavior that sets network of natural numbers apart from other complex networks. In the network presented here, each node has an identity which is the number attached to it and this defines a natural order on the nodes. This means that we can study various properties of nodes as a function of their labels. This is not possible for other networks because no such unique labeling exists for the nodes. Here we specifically consider local clustering coefficient of nodes and study its behavior as a function of node index. We find that the clustering coefficient $c_{i}$ of node $i$ varies seemingly irregularly. However, when $c_{i}$ is plotted against $i$, a global pattern is seen. In Fig.~\ref{cluster_stretch} we show this pattern for three different network sizes. For better visualization, the plots are shown only for relatively small network sizes.
From the figure, it is clear that the global pattern of the local clustering coefficient gets stretched as the size of the network increases such that the nature of the pattern remains the same. We call this new kind of similarity as ``stretching similarity" and this seems to be a unique feature of this network, not so far reported for any other complex network. We note from plots in Fig.~\ref{cluster_stretch} that for a network with size $N$  some discontinuous vertical steps occur approximately at values $N/2,N/3,N/4,\cdots$. Also, we observe a band of numbers with clustering coefficient $1$ between $N/3$ and $N/2$ and these numbers correspond to prime numbers and their powers in that range. This can be seen by the following argument. Consider any prime number $p$ in the interval $(N/3,N/2)$. On the lower side, it is connected only to $1$ while on the upper side, it would be connected only to its multiples. However, all the numbers in this range would have only one multiple $2p$ up to $N$. Thus, three numbers $1,p,2p$ form a triangle and hence clustering coefficient of number $p$ must be $1$. A similar argument for prime powers in this range tells that they also have clustering coefficient $1$. There is another band of numbers with clustering coefficient exactly $0$ between $N/2$ and $N$ which are also prime numbers. This is because all the primes in this range are connected only to $1$ making their clustering $0$.

Now we discuss the local clustering coefficient for the composite numbers between $N/2$ and $N$. For a vertex $n$, the only neighbors are the proper 
divisors of $n$ i.e. $m$ such that $1 \leq m < n$ and $n$ is divisible by $m$.

Let $n = \prod \limits_{i=1}^{k} p_{i}^{j_{i}}$ be the  factorization of $n$ as the product of (distinct) prime powers. The fundamental theorem of arithmetic states that such a factorization is unique up to a reordering of the primes $p_{i}$'s. It can be observed that every divisor $m$ of $n$ is of the form 
$m = \prod \limits_{i=1}^{k} p_{i}^{\ell_{i}}$ where 
$0 \leq \ell_i \leq j_i$ for every $ 1 \leq i \leq k$. 

Any two neighbors $m = \prod \limits_{i=1}^{k} p_{i}^{\ell_{i}}$ and $m' = \prod \limits_{i=1}^{k} p_{i}^{\ell^{'}_{i}}$ such that $m < m'$ are adjacent to each other if and only if $\ell_{i} \leq
\ell^{'}_{i} $ for all $i$.

 Thus the clustering coefficient of $n$ in the network of size $N$ is given by,
\begin{equation}
c_{n} = \binom{s-1}{2}^{-1} \left[\left(\sum \limits_{\ell_{1} = 0}^{j_{1}} \sum \limits_{\ell_{2} = 0}^{j_{2}} \cdots \sum \limits_{\ell_{k} = 0}^{j_{k}} [(\ell_1+1) (\ell_2+1) \cdots (\ell_k+1) - 1] \right) - \left[s-1 \right] \right].
\end{equation}

where $s = (j_1 + 1) (j_2 + 1) \cdots (j_k + 1)$. 

\begin{equation}
\therefore ~ c_{n} = \binom{s-1}{2}^{-1}\left(\prod\limits_{i=1}^{k}\binom{j_{i}+2}{2}-2s+1\right)
\end{equation}

From the above expression it follows that value of $c_{n}$ depends only on the number of distinct prime factors of $n$ and the powers $j_i$'s which appear in the prime factorization of $n$; but not on the actual primes which appear there. Thus for any given $j_1, j_2, \ldots j_k$, the value $c_{n}$ is constant for every $n$ in the range $N/2 < n \leq N$ such that  $n = \prod \limits_{i=1}^{k} p_{i}^{j_{i}}$ for some set of $k$ distinct primes $p_1, p_2, \cdots p_k$. This explains the occurrence of horizontal dotted lines in the plot for local clustering coefficients. 

Similarly, the clustering coefficients for other $n$ can be computed and it can be observed that they depend on the powers and the number of distinct prime factors of $n$ as well as the range   in which $n$ belongs that is $r$ such that $N/(r+1) < n \leq N/r$. Here one has to also consider the number of multiples of $n$ in the range $1,2,\cdots, N$. This leads to possibly different values of clustering coefficients. This explains the occurrence of demarked regions like $N/2$ to $N$, $N/3$ to $N/2$, $N/4$ to $N/3$ etc in the plot for local clustering coefficients. For any $N$ there will be sufficient number of primes in the range $[1,N/2]$ and choices for $j_{i}$ such that the pattern of horizontal lines between $N/2$ to $N$ remains the same. Also, the demarked regions have similar structures. This provides a possible explanation for the observed stretching similarity in the clustering coefficients as $N$ is changed (Fig.~\ref{cluster_stretch}).

\begin{figure}[ht]
\begin{center}
\includegraphics[width=1\columnwidth]{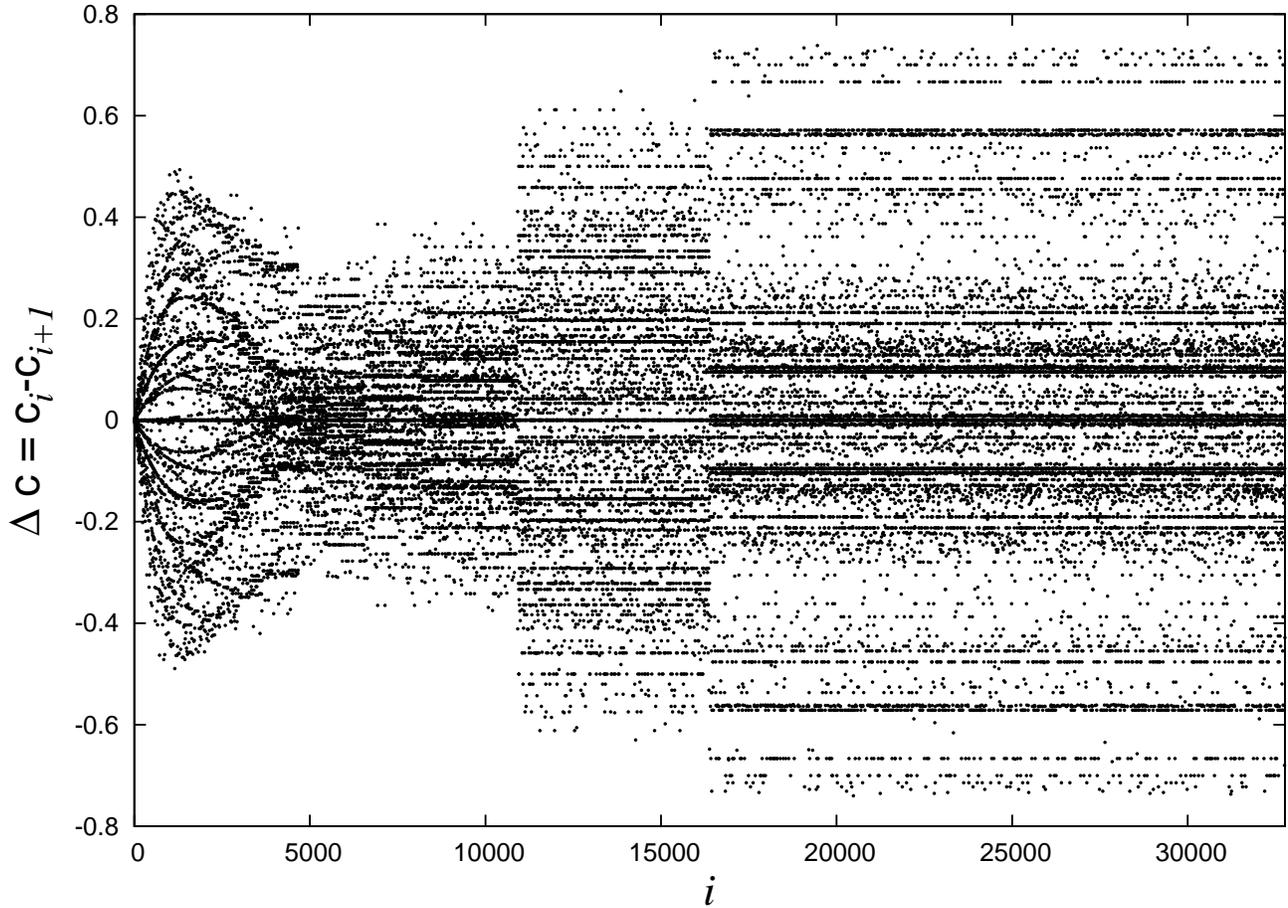}
\caption{\label{cluster_diff} \textbf{Difference between clustering coefficients of successive nodes $i$ and $i+1$ as a function of index $i$}. This pattern is symmetric about the line $\triangle c = 0$ and also shows stretching similarity.}
\end{center}
\end{figure}

We also observe an interesting pattern when we plot the difference $\triangle c = c_{i}-c_{i+1}$ as a function of $i$ in Fig.~\ref{cluster_diff}. We find that this pattern is symmetric about $\triangle c = 0$ which can be quantified by finding the local density of values in the plot (see Methods). With increasing size of the network, this pattern also shows stretching similarity.
\textbf{Topological characteristics of the network.} In the present section, we discuss how three of the most important quantities average degree, global clustering and assortativity coefficient vary with the size of the network. 

\subsection*{Average degree} Here we derive an approximate expression for the average degree of the network as a function of its size. By definition, the average degree of the network is given by: 

\begin{equation}
\label{ave_deg}
<k>_{n} = \frac{2m}{n}
\end{equation}
\begin{figure}[ht]
\begin{center}
\includegraphics[width=1\columnwidth]{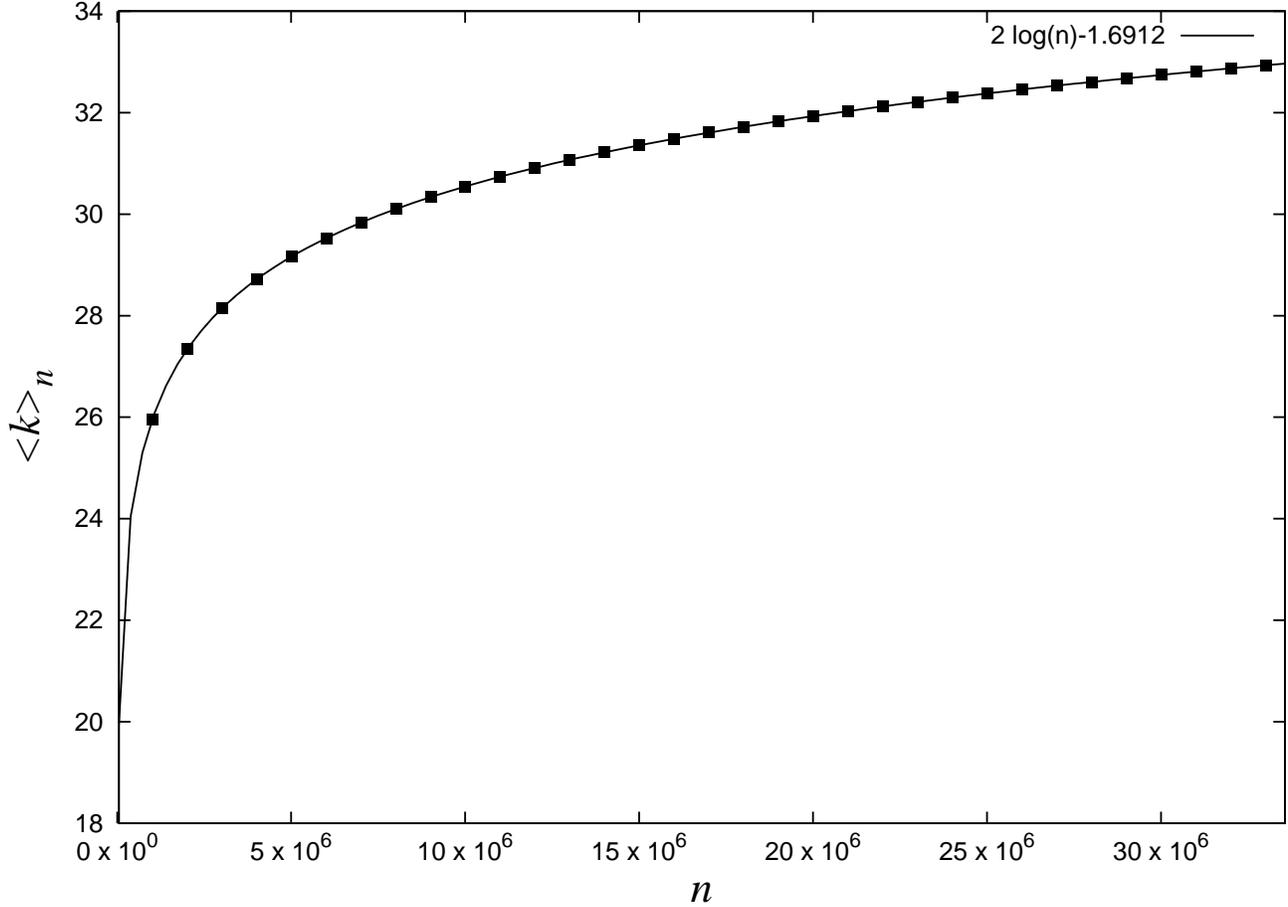}
\caption{\label{ave_deg_variation} \textbf{Average degree of the network as a function of size.} The solid dots represent the actual values calculated by direct numerical simulations while the solid line is plotted using the analytic expression (\ref{final_ave_deg}).}
\end{center}
\end{figure}
where $m$ is the total number of edges in the network and $n$ is the size of the network. The value of $m$ is also equal to the sum of the elements in lower (or upper) triangular part of the adjacency matrix. To find this sum, we interpret the second index of element $A_{ij}$ of adjacency matrix to be the divisor of first index if $A_{ij}=1$. In other words, let $A_{ij} = 1$ if and only if $i>j$ and $j|i$. Then the sum of the elements in the lower triangular part of the matrix is equal to the number of integers of the form $kj$ with $k\geq 2$ and $kj\leq n$. However, whenever $j>\frac{n}{2}$ all the entries in the in the $j^{th}$ column of the lower triangular part of $A$ are zero. Let $\lfloor{x}\rfloor$ denote the greatest integer $\leq x$. Then $m$ is given by:

\begin{equation}
\label{m_floor}
\begin{aligned}
m &= \sum\limits_{j=1}^{n/2}\left(\left\lfloor{\frac{n}{j}}\right\rfloor -1\right)\\
&= \sum\limits_{j=1}^{n}\left\lfloor{\frac{n}{j}}\right\rfloor-\sum\limits_{n/2<j\leq n}\left\lfloor{\frac{n}{j}}\right\rfloor-\frac{n}{2}
\end{aligned}
\end{equation}

It is well known that the first term on the right satisfies an estimate as follows \cite{Apostol}:
\begin{equation}
\label{Dirichlet}
\sum\limits_{j=1}^{n}\left\lfloor{\frac{n}{j}}\right\rfloor = n\ln n + n(2\gamma-1) + O(\sqrt{n})
\end{equation}
where $\gamma$ is Euler-Mascheroni constant. Also we  observe that:
\begin{equation}
\label{floor}
\left\lfloor{\frac{n}{j}}\right\rfloor = 1 \quad \forall~ \frac{n}{2}<j\leq n
\end{equation}

From Eqs.(\ref{ave_deg}),(\ref{m_floor}),(\ref{Dirichlet}),(\ref{floor}), it follows that:

\begin{equation}
<k>_{n} ~=~ 2\ln n + 2(2\gamma-1)-2+O(\frac{1}{\sqrt{n}}) \quad \text{as} ~n\rightarrow\infty
\end{equation}

Since $\gamma\approx 0.5772$, in the limit of large $n$, we get,

\begin{equation}
\label{final_ave_deg}
<k>_{n} ~\sim~ 2\ln n - 1.6912
\end{equation}

This means that the average degree of the network increases logarithmically with the size and this variation is plotted in Fig.~\ref{ave_deg_variation} (solid line) using Eq.(\ref{final_ave_deg}). We calculate this numerically by growing the network up to $N = 2^{25}$ and the results obtained, shown by solid dots in Fig.~\ref{ave_deg_variation}, are found to agree exactly with analytic expression (\ref{final_ave_deg}). Since the average degree of the network increases with size, the degree distribution of the network is not stationary though as shown in the previous section, the network is scale-free at each stage (see Methods). 

\subsection*{Global clustering coefficient}
The global clustering coefficient of the network quantifies the density of closed triplets in the network. A connected triplet in the network is the set of $3$ nodes connected to each other with exactly $2$ links. A closed triplet is the set of $3$ nodes connected to each other with exactly $3$ links. A triangle in the network counts as three closed triplets (one centered at each node of the triangle). The global clustering coefficient of the network is then defined as:
\begin{equation}
\label{global_cluster}
C = \frac{3\times \text{Number of triangles}}{\text{Number of connected triplets}}
\end{equation}
We estimate the number of triangles $T_{n}$ in the network using the following strategy. Let us fix a vertex $i$ and calculate the number of triangles in which $i$ is the smallest vertex. The number $i$ has $\left\lfloor{\frac{n}{i}}\right\rfloor-1$ proper multiples in the range $[1,n]$. Each of them is of the form $ki$ where $k=2,3,...,\left\lfloor{\frac{n}{i}}\right\rfloor$. Thus, $T_{n}$ is given by:

\begin{equation}
T_{n} = \sum\limits_{i=1}^{n}\sum\limits_{k=2}^{\left\lfloor{\frac{n}{i}}\right\rfloor}\left\lfloor{\frac{n}{ki}}\right\rfloor 
\end{equation}
\begin{figure}[ht]
\begin{center}
\includegraphics[width=1\columnwidth]{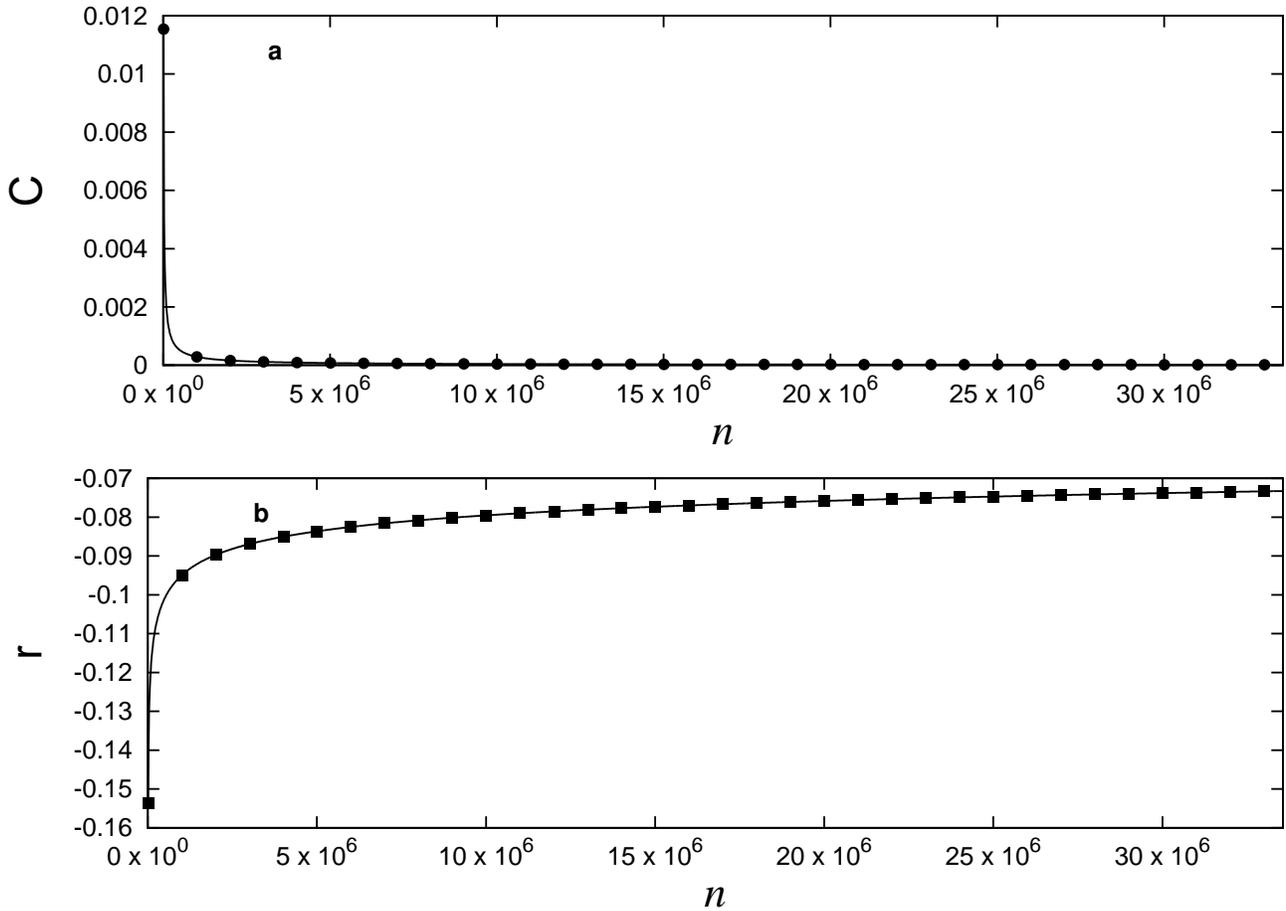}
\caption{\label{cluster_assort_vs_size} \textbf{Global clustering coefficient and assortativity coefficient as a function of size of the network.} ({\bf a}) The global clustering coefficient (see Eq.(\ref{global_cluster})) decays to $0$ as the size of network increases. ({\bf b}) The assortativity coefficient $r$ (see Eq.(\ref{assort})) also seems to reach $0$ asymptotically though it always remains negative.}
\end{center}
\end{figure}
Using the integral approximation for the above:
\begin{equation}
T_{n} \sim n \sum \limits_{i=1}^{n}  \int \limits_{x=2}^{n/i} \frac{1}{ix} dx \sim n \sum \limits_{i=1}^{n} \frac{1}{i} \left( \ln \frac{n}{i} - A \right)
\end{equation}
The above is bounded by,
\begin{equation}
n \sum \limits_{i=1}^{n} \frac{1}{i} \left( \sqrt{\frac{n}{i}} - A \right) \sim n\sqrt{n}\int\limits_{x=1}^{n}\frac{dx}{x^{3/2}}- An\int\limits_{x=1}^{n}\frac{dx}{x}\sim Bn-An\ln n
\end{equation}
Here $A$ and $B$ are constants. Hence we see that:
\begin{equation}
\label{triangles_order}
T_{n} \leq O(n) + O(n \ln n) + o(n^2)
\end{equation}

In particular, 
\begin{equation}
\label{T_n}
T_{n} = o(n^2)
\end{equation}

Let $U(n)$ be the the number of connected triplets in the network after $n^{th}$ stage. Then $U(n)$ is given by:

\begin{equation}
U_{n} = \sum\limits_{i=1}^{n}(k_{i}^{2}-k_{i}) =  n\left\langle k^{2}\right\rangle-n\left\langle k\right\rangle \sim O(n\left\langle k^{2}\right\rangle)
\end{equation}

Since we have (Fig.~\ref{deg_distri}) observed that the degree distribution of the network follows a power law $k^{-\alpha}$ with $\alpha \sim 2$, we see that the proportion $p(k)$ of nodes with degree $k$ is $\sim k^{-2}$.

Thus, the expectation of the variable $k^2$ satisfies: 
\[ \left\langle k^2 \right\rangle = \sum \limits_{k=1}^{n} k^2 p(k) \sim n \]

Hence we see that 
\begin{equation}
\label{U_n}
U_{n} \sim n \left\langle k^2 \right\rangle = O(n^2)
\end{equation}

From Eqs.(\ref{global_cluster}), (\ref{triangles_order}) and (\ref{U_n}), the global clustering coefficient decays to zero as the network size goes to infinity. We verify this by numerically computing the global clustering coefficient and this is shown in Fig.~\ref{cluster_assort_vs_size}.{\bf a}. However, we note that the Watts-Strogatz clustering coefficient $C_{WS}$ of the network (which is defined as the average of all local clustering coefficients over all the nodes of the network \cite{Watts_1998}) does not decay to zero and instead reaches to a constant value $\sim 0.6$. This is clear from Fig.~\ref{cluster_stretch} since the pattern repeats with stretching similarity as the network size increases. To the best of our knowledge, there is no other network in which $C_{WS}$ saturates to a high non-zero value but the global clustering coefficient decays to $0$.

\subsection*{Assortativity coefficient}
The correlation of degrees in the network is an important quantifier of the network structure \cite{Newman_2002}. If in a network the high degree nodes tend to connect to low degree nodes (i.e. if the network has negative degree correlations), then the network is said to be dissortative in structure whereas if similar degree nodes tend to connect to each other, network is said to be assortative. All the real networks except social networks are dissortative \cite{Newman_2002} and this has been explained using the fact that the dissortative state is the most likely state of scale-free networks \cite{Johnson_2010}. The assortative/dissortative nature of networks can be quantified using the assortativity coefficient \cite{Newman_book}:

\begin{equation}
\label{assort}
r = \frac{\sum_{ij}(A_{ij}-k_{i}k_{j}/2m)k_{i}k_{j}}{\sum_{ij}(k_{i}\delta_{ij}-k_{i}k_{j}/2m)k_{i}k_{j}}
\end{equation}
where $k_{i}$ is the degree of the $i^{th}$ node, $A_{ij}$ is the $(i,j)^{th}$ element of the adjacency matrix, $m$ is the total number of edges in the network and $\delta_{ij}$ is the Kronecker delta.

In Fig.~\ref{cluster_assort_vs_size}.{\bf b} we show the dependence of $r$ on the size of the network and in spite of irregularity in the divisibility pattern, $r$ has a smooth behavior with $n$. It can be seen that $r$ always remains negative though asymptotically it seems to reach the value $0$ implying that the network is dissortative. The dissortative nature of the network of natural numbers is understandable from the following argument. For any link in this network, the one end of the link is divisor (node $A$) and other is multiple (node $B$). Hence node $A$ is also connected to all the nodes which are multiples of $B$ but the reverse is not true. This means that the degree of node $A$ always tends to be very high as compared to degree of node $B$ for a given size of the network giving the negative value for the overall correlation coefficient. 

We also find that all the important statistical properties of the network like stretching similarity, degree distribution, clustering-degree correlation etc. are very robust to the removal of even the biggest hubs like numbers $1,2,3,..$. This shows that the global divisibility pattern of natural numbers does not depend only on the few nodes but instead is built by contributions from all the nodes. (See Methods)

\section*{Discussion}
The network of natural numbers constructed using divisibility relations looks like real networks in many characteristics like degree distribution, clustering and degree correlations. We show how insights into the divisibility patterns of natural numbers can be obtained using the framework of complex networks, where we consider both composite and prime numbers in a single undirected network with links generated using both multiples and divisors. Some of the interesting results that we get are the scale-free nature of the network with a non-stationary distribution and the existence of stretching similarity. We validate the existence of power-law in the distribution and estimate the corresponding power-law index using rigorous techniques from statistical inference advocated by Clauset et al \cite{Clauset_2009}. We find that the average degree of the network grows logarithmically with the size of the network and we find the exact formula for its behaviour analytically. We also find that the global clustering coefficient of the network reaches to the value  $0$ while the average clustering coefficient $C_{WS}$ saturates to a high value. All these results are validated by extensive numerical calculations for network up to size $2^{25}$.

We also find that there exists a pattern in the local clustering coefficients that reflects universality in the organization of natural numbers in terms of their prime constituents. We observe that this pattern has a stretching similarity which is a reflection of the nature of prime factorization of natural numbers. Also, the behavior of characteristics like average degree, global clustering and assortativity coefficients for this network vary quite smoothly and hence may help us to understand better the divisibility relations between natural numbers. In conclusion, the work presented here describes an interesting perspective on the divisibility relations of natural numbers and has potential to become an important tool in the investigation of the properties of natural numbers.

\section*{Methods}
\textbf{Establishing the scale-free nature of the network.} The shape of the degree distribution of the network in Fig.~\ref{deg_distri} hints at the existence of asymptotic power law in the distribution ($p(k)\sim k^{-\alpha}$ for $k\ge k_{min}$). However a visual inspection to find $k_{min}$ and least square fit and related methods to find the exponent $\alpha$ of the power law are known to produce very bad estimates \cite{Goldstein_2004}. Hence we use the method of maximum likelihood for the degree sequence of the network to find scaling index $\alpha$ of the power-law distribution \cite{Clauset_2009}. For this, we initially assume that the sequence is drawn from a distribution that follows a power law $k^{-\alpha}$ for all $k$ after $k\geq k_{min}$. To find this $k_{min}$, we use the approach proposed by Clauset et al. \cite{Clauset_2007}. The idea behind this method is to choose that value of $k$ as $k_{min}$ which makes the probability distribution of the data and best-fit power-law model as similar as possible above $k_{min}$ where we use Kolmogorov-Smirnov statistic as the distance between two distributions \cite{Clauset_2009}. After finding $k_{min}$ using this method, the best estimation for scaling exponent $\alpha$ is given by:

\begin{equation}
\label{alpha_eqn}
 \alpha = 1 + N\left[\sum\limits_{i=1}^{N}\ln\frac{k_{i}}{k_{i}-\frac{1}{2}}\right]^{-1}
\end{equation}

where $k_{i}$, $i=1,\cdots,N$ are values of $k$ such that $k_{i}\geq k_{min}$. For the network of size $2^{15}$, the value $\alpha$ is obtained here as $\sim 2$.

To validate the existence of power law, we use the approach described in Clauset et al\cite{Clauset_2009}. In this approach we generate many synthetic data sets from a true power-law distribution and measure how far they fluctuate from the power-law type of behavior. We then compare the results of similar measurements on the observed data. If the observed data set is much further from the power-law form than the synthetic one, the power-law is rejected. The $p-$value is defined as the fraction of the synthetic distances that are larger than the empirical distance. A large $p-$value is indicative of existence of power law in the data. In the present work we calculate the $p-$values for three different sizes of the network: $N = 256, 512, 1024$. For this, we generate $2500$ synthetic data sets which gives $p-$values accurate up to two decimal places as $0.62$, $0.95$ and $0.98$ respectively. The existence of power-law degree distribution for this network is thus confirmed by the fact that $p-$values rapidly converge to $1$ as the network size increases. 

The distribution in Fig.~\ref{deg_distri} is plotted with logarithmic binning with the successive bin sizes equal to successive powers of $2$ and the count in each bin is normalized by dividing the count by the bin-width. The same strategy is used to show the dependence of local clustering coefficient $c(k)$ on degree $k$ in Fig.~\ref{cluster_vs_deg}.

\begin{figure}[ht]
\begin{center}
\includegraphics[width=1\columnwidth]{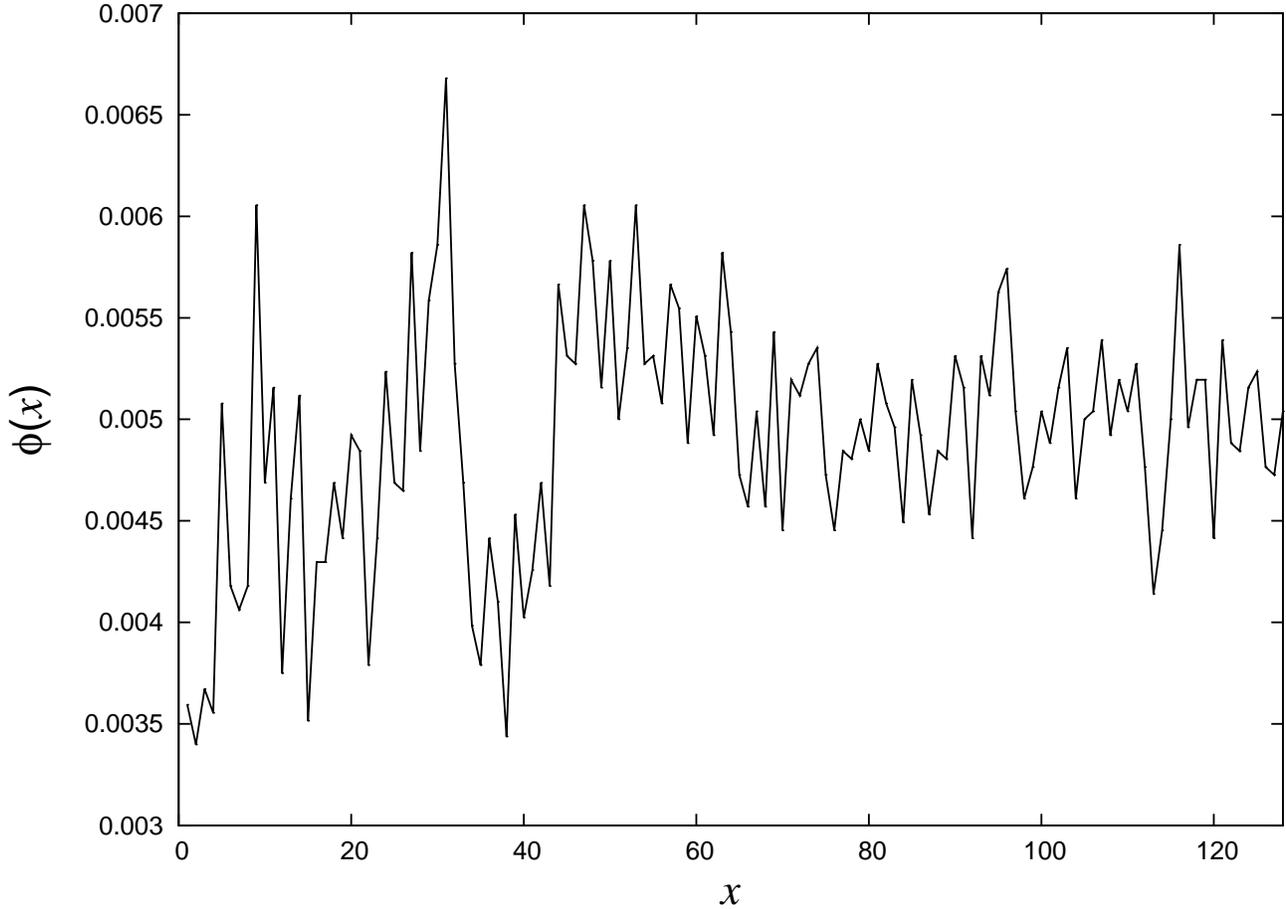}
\caption{\label{sym_quant} \textbf{Symmetry quantifier for the Fig.~\ref{cluster_diff} as given by Eq.(\ref{density_diff}).} The values of $\phi$ are very close to zero for all horizontal pixel indices establishing the approximate symmetry for the pattern.}
\end{center}
\end{figure}

\begin{figure}
\begin{center}
\includegraphics[width=1\columnwidth]{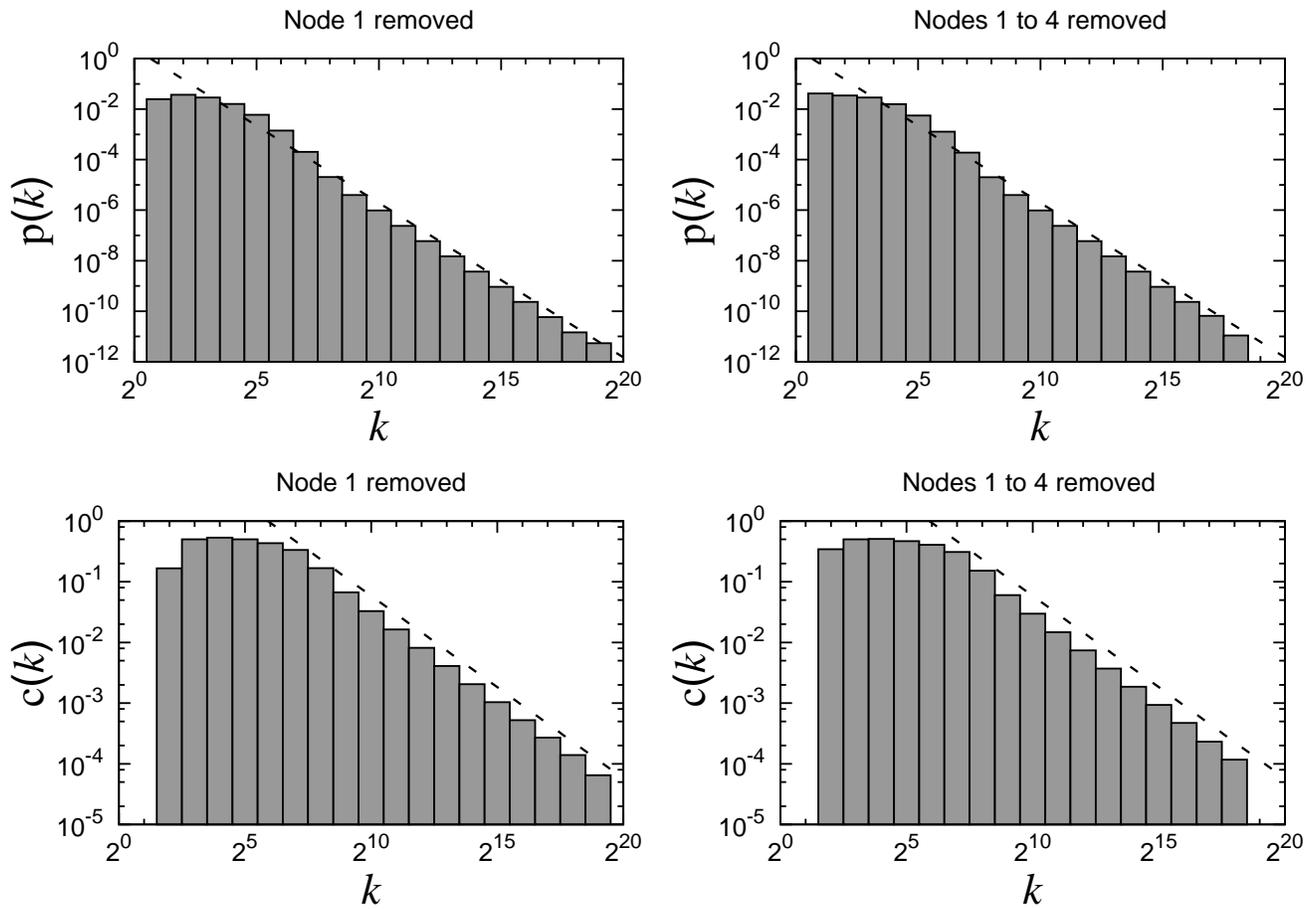}
\caption{\label{Removal} The degree distributions of the network of natural numbers after removing nodes from $1$ to $4$. The distributions follow a power-law similar to the original network.}
\end{center}
\end{figure}

\textbf{Symmetry in difference of successive local clustering coefficients.} To establish the global symmetry of difference in local clustering values $\triangle c$ around the horizontal axis $\triangle c = 0$ (Fig.~\ref{cluster_diff}) for any value of $N$, we calculate the local density of points in the plot. For this, we divide the horizontal axis into $2^7=128$ cells and vertical axis into $200$ cells. The whole plot then gets divided into pixels of dimension $0.01\times 2^{N-7}$. We define density $\rho(x,y)$ of a particular pixel $(x,y)$ as the ratio of the number of points present in the pixel to the maximum number that can be there which is equal to $2^{N-7}$ (all the points on y-axis with difference less than $0.01$ are to be considered same so the vertical dimension of each pixel is just $1$). For each $x$ we calculate the absolute difference between the corresponding pixels on each side of the line $\triangle c = 0$. If the pattern is symmetric then these absolute differences are expected to be small. We calculate the average of such differences as: 

\begin{equation}
\label{density_diff}
\phi(x) = \frac{1}{100}\sum\limits_{y=1}^{100}|\rho(x,y)-\rho(x,-y)|
\end{equation}
In Fig.~\ref{sym_quant} we show $\phi(x)$ as a function of $x$ and as is clear from the figure, all $\phi$ values are very close to $0$ confirming that the pattern is indeed symmetric.
\textbf{Removal of hubs from the network.} To test the robustness of the various statistical properties of the network against the removal of hubs from the network, we simulated the network of natural numbers removing numbers $1$ to $4$ step by step. When number $1$ is removed from the network, all the prime numbers between $N/2$ and $N$ become isolated and these remain as the only isolated nodes. This means that in this case the network consists of a giant component along with many isolated nodes. We find that such a removal does not affect the degree distribution and clustering-degree correlation too much and qualitatively the network remains scale-free with the same power-law index as for the original network. The other properties like average degree, clustering coefficients and assortativity do change to some extent by this removal but qualitatively remain the same. The  plot of degree distributions after removing hubs is shown in Fig.~\ref{Removal}.
\section*{Acknowledgements}
S.M.S. is supported by Senior Research Fellowship from University Grants Commission, Delhi, India. C.B. is supported by DST-INSPIRE faculty scheme, award number [IFA-11MA-05]. Authors acknowledge Joel Ornstein for making the python implementations of some of the methods used in this paper available to us.

\section*{Author contributions statement}
S.M.S. proposed the idea and performed the numerical simulations. C.B. derived the results analytically. G.A. supervised the study. All authors discussed the results and prepared the manuscript.

\section*{Additional information}
The authors declare no competing financial interests.

\end{document}